# Emulating short-term synaptic dynamics with memristive devices

Radu Berdan, Eleni Vasilaki, Ali Khiat, Giacomo Indiveri, Alexandru Serb, Themistoklis Prodromakis

**Neuromorphic architectures offer great promise for achieving computation capacities beyond conventional Von Neumann machines. The essential elements for achieving this vision are highly scalable synaptic mimics that do not undermine biological fidelity. Here we demonstrate that single solid-state $TiO_2$ memristors can exhibit non-associative plasticity phenomena observed in biological synapses, supported by their metastable memory state transition properties. We show that, contrary to conventional uses of solid-state memory, the existence of rate-limiting volatility is a key feature for capturing short-term synaptic dynamics. We also show how the temporal dynamics of our prototypes can be exploited to implement spatio-temporal computation, demonstrating the memristors full potential for building biophysically realistic neural processing systems.**

Connections among neurons are more than simple cables that transmit signals: they are known to exhibit temporal, revertible dynamics in short time scales. These features are termed short-term plasticity (STP) and are well described both experimentally[1-7] and in the context of specific models[8-10]. The role of STP in neuronal computation is thought to be related to temporal processing, see for instance[11] or the work by Carvalho and Buonomano[12], where STP was shown to enhance the discrimination ability of a single neuron, i.e. a tempotron[13], when presented with forward and reverse patterns. Synapses with short-term plasticity are also optimal estimators of presynaptic membrane potentials[14], and correlate to specific brain connectivity configurations[15,16], that are hypothesized to emerge via learning processes[17-19]. Most interestingly, the vast majority of artificial and neuromorphic brain-like systems focus on stable modifications of connections, known as long term plasticity, which are assumed to be the basis of memory, and do not make use of the computational power that short-term plasticity may provide, but rather demonstrate a behavior akin to short term dynamics. Moreover, they largely ignore the fact that synapses are inherently unreliable and there is often a large variance in their response to a specific signal, also apparent in short-term dynamics.

Among the several candidates for fabricating brain-like, neuromorphic systems, memristors[20] are particularly promising: their characteristic signature of hysteresis is typically noticed in systems and devices that possess certain inertia, manifesting memory, including neural systems. Notwithstanding the several efforts for implementing these mechanisms via Complementary Metal-Oxide-Semiconductor (CMOS) topologies and emerging nanoscale cells, that were able to capture short-term plasticity[21-24], memristors have the potential to alleviate challenges imposed by CMOS implementations[25]. They can reduce energy consumption and size by exploiting their simple (two terminal) architecture and small footprint[26], their capacity to store multiple bits of information per cell[27] and the miniscule energy required to write distinct states[28]. Memristors have been shown to exhibit properties akin to long-term plasticity, such as Spike-timing Dependent Plasticity (STDP)[29], along with STDP variations[30,31], in compact and large scale cross-bar architectures[32,33]. The majority of such approaches relies on non-volatile memory-state transitions based upon phase-change[31,34] mechanisms or the diffusion of ionic-species within an active core[35-37].

The functional properties of memristive devices are however associated with irreversible rate-limiting electro/thermo-dynamic changes that quite often bring them in far from equilibrium conditions, rendering a rate-limiting volatility[38]. While the majority of researchers focus in demonstrating how non-volatile conductance changes in memristive devices can resemble the STDP induced changes in real synapses, only few studies exist that leverage metastable effects in memristive devices[39] for reproducing short-term synaptic dynamics[40]. However in most cases the equivalence between the physics of memristive devices and the physics governing the behavior of real synapses has been shown only at an abstract qualitative level. Here we show how single $TiO_2$ memristors are capable of capturing short-term synaptic dynamics using the same experimental protocols and models used to validate the response of real synapses. We also show for the first time how the short-term dynamic properties of our memristive devices can be exploited for implementing spatio-temporal computation, following biologically realistic neural computation paradigms[12,13,41].

## Switching dynamics of $TiO_2$ memristors

Hysteresis is typically noticed in elements that possess certain inertia; manifesting memory[42]. Verily, considerably larger systems are also known to have similar non-linear signatures[43]. Particularly in nanoscale memristors this inertia is ascribed to Joule heating[44], the electrochemical migration of oxygen ions[45] and vacancies[46,47], the lowering of Schottky barrier heights at interfacial states[48], the phase-change[49,50] and the formation/rupture of conductive filaments[51,52] in a device's active core. Excellent reviews that cover distinct switching mechanisms exist[27,53,54], which overall can be classified in three categories[55]: (1) displacement of ionic species, (2) phase-change and (3) the formation of conductive filaments. Recently we demonstrated that substantial resistive switching is only viable through the formation and annihilation of continuous conductive percolation channels[56] that extend across the whole active region of a device, no matter what the underlying physical mechanism is. Innately, in the case of $TiO_2$ memristors the underlying functional mechanism is a manifestation of all three categories: ionic-species ($O^{-2}$ vacancies[47] and/or $Ti^{n+}$ interstitials[57]) are translocated within the active core that cause the formation of locally reduced Magnelli phases[51], which in turn extend along the $TiO_2$ core as current percolating branches that can be annihilated via Joule heating[54].

Our memristive device qualitatively represents a synapse (inset I of Figure 1a), with its conductance corresponding to the notion of a synaptic efficacy modulated via the arrival of a spike, i.e. a pulse applied pre-synaptically to the device's top electrode (TE), shown in inset II of Figure 1a. The post-synaptic current entering the artificial neuron, from the device's bottom electrode (BE), is proportional to the memristive conductance. Figure 1a depicts a microphotograph of one of our fabricated crossbar type $TiO_2$-based memristors (fabrication details are given in Methods). The device comprises two Pt electrodes (TE and BE) that are separated by a stoichiometric $TiO_2$ active core (cross-section is shown in inset II of Figure 1a). Following an electroforming step (depicted in Figure S1), the devices' electrical characteristics were first investigated via positive/negative ±2V voltage sweeps, resulting into a bipolar mode of switching: positive sweeps cause low- (LRS) to high-resistive state (HRS) transitions, while negative ones cause HRS to LRS transitions. The corresponding current-voltage (I-V) characteristics, with the classical pinched-hysteresis memristor signature is shown in the supplementary material Figure S2a.

In order to induce a non-volatile (stable) resistive transition, a critical energy barrier $E_i$ has to be exceeded that will allow toggling between long-term thermodynamically stable (non-volatile) states. It is interesting to note that the required activation energy $E_i$ for toggling between stable states depends upon the previous state of the device, causing it to act as a non-linear accumulator. This is demonstrated in Figure S2c by employing subsequent identical voltage pulses that result into a non-uniform modulation of the effective resistance of our prototypes. Particularly, the notation used in Figure 1b denotes that a non-uniform lowering[38] (increasing) of the barrier occurs as the applied electric field elicits a HRS to LRS (LRS to HRS) non-volatile transition. In the case however the energy provided to the system is less than the corresponding activation energy $E_i$, a transient (volatile) response can be temporally induced, with the initial equilibrium state being eventually restored, as depicted in Figure 1c. Such a response is equivalent to the revertible, use-dependent modifications observed in synaptic connections, known as "short-term" plasticity (STP). This state volatility occurs in memristors, due to metastable phase-transitions within the functional active core and precedes the induction of any long-term phase-change in the device's bulk via forming (field-driven) or annihilating (thermally) stable conductive percolation channels. The kinetics of this process are concurrently governed by mass diffusion[29] and nucleation[58] processes that also form the basis of STP events in the atomic switch[39]. The accumulating nature of these events elicit activity-dependent state modulations that can cause unpredictable switching trends, which in combination with the exhibited volatility add a substantial probabilistic component in the devices' switching dynamics[59]. Here, we employ this inherent rate limiting volatility of our TiO₂-based memristive devices for emulating the temporal (short-term) behavior of real synapses. In what follows, we utilise our memristive synapses with an Integrate-and-Fire (IF) neuron to practically exploit short-term plasticity for spatiotemporal computation.

### Non-associative, short-term synaptic plasticity in single TiO₂ memristors

The transient conductance response of the same TiO₂ memristor is first utilized to model STP changes in synapses, as shown on Figures 2a and 2b. Figure 2c shows the voltage pulse pattern used to produce these responses (see supplementary material Figure S3 for a detailed description of the experimental setup). Each voltage pulse induces an increase in the conductance of the device, which then tends to slowly decay to its original state. Subsequent pulses have a similar effect that however depend on the previous resistive state of the device, the conductance peaks can be lower or higher in magnitude than the first peak, similar to short term plasticity mechanisms, as observed in biological synapses[40].

In particular, the form of short-term plasticity emulated with this device is "short-term facilitation". Figures 2a and 2b show two different cases of facilitation: in Figure 2a we reproduce the classical form of short-term facilitation (here denoted as STP-F), where each input pulse has the effect of increasing the conductance, including its peak response. At the arrival of the first pulse the memristive conductance increases from the base line (blue line) allowing current to pass through to the neuron, causing an increase on the membrane potential of the connected neuron, shown by the black solid line. On the second spike, the conductance transiently increases even more and so at the third spike, leading to a behavior akin to short-term potentiation, where the synaptic efficacy transiently changes in short time scales. The observed phenomenon is clearly non-linear and cannot be explained by the linear summation of the input signals. We have deliberately chosen a fast membrane potential time constant so that no residuals are remaining from the previous spikes, demonstrating that each spike indeed contributes to the membrane voltage by a different amount. Modeling the measured conductance peaks by a reduced version of the Tsodyks-Markram model[60] (see supplementary section), verified quantitatively the equivalence with biological synapses and also revealed that the time constants involved in the process are close to typical biological values. At a subset of our experiments, particularly when the initial conductance of the material is a relatively higher state, the same protocol will lead to a transient conductance at the second and third spike clearly lower than the initial one. We call this phenomenon saturation (STP-S), and we provide a hypothesis of how this deviation from the typically observed response of the memristors occurs. We argue that with every pulse, the potential alignment of filaments comprising of reduced $TiO_2$ leads to a higher conductive state, following the model presented in Figure 1b. In the case where the provided energy does not exceed the instantaneous $E_i$ that would cause the device to undergo LTP or LTD (in the case of a significant energetic overshoot), such a transition is revertible and after a transient the conductance returns to its initial state. However, if the provided energy will be (is about to be) dissipated via an existing (partially formed) filament, at the same time annihilating it, we will most likely observe saturation in the response, as shown in Figure 2b. We argue that the exhibited STP-S response stems from the mobility saturation of the available ionic resources (mainly $O^{-2}$ vacancies) in the vicinity of a partially reduced $TiO_{2-x}$ volume, i.e. a partially formed filament, particularly when the device has been previously stimulated. Clearly, a finite number of mobile resources exist within the volume of interest that can play a role in reducing $TiO_2$ from insulating towards (semi-) metallic phases, i.e. towards forming a conductive filament. And as single devices could in principle host multiple filaments[46] within their functional cores, alike short-term plasticity phenomena can be triggered across a wide conductance spectrum (see Figure S8 in supplementary material).

In a concurrent experiment, the same device was subjected to a train of 3 consecutive voltage pulses of -4V, 10µs wide and inter-pulse interval $t_{int}$=400ms. This sequence was repeated 600 times with a recovery interval between sequences $t_{rec}$=10s, to allow for the device's state being restored. The pre-stimuli initial conductance was found to vary within 2.85 and 3.1 µS. This range was divided into 17 equal conductance bins with STP-F and STP-S events discriminated (as in Figures 2a and 2b) and plotted with respect to the device's initial conductance (Figure S9). It is interesting to note that during the experiment the initial conductance range increased to values above 2.95 µS, possibly due to the partial formation of a new stable filament. This effect yielded a new equilibrium conductance at which the device could settle. Nonetheless, for both stable-state conditions, STP-S events are more likely to manifest at higher conductance levels than STP-F events, as shown by the corresponding probabilities of STP-F (Figure 2d) and STP-S (Figure 2e) occurrences. This illustrates the strong probabilistic switching nature that under the classical ReRAM context will contribute substantially to the devices unreliability.

In relation to the occurrence of the STP events presented in Figures 2d,e, we have recorded all events of a single device when repeatedly excited with a stimulating scheme comprising three voltage pulses of 4V, 10µs wide, $t_{int}$=200ms and $t_{rec}$=20s, as illustrated in Figure 2f. Facilitating (depressing-like) events have been colored mapped with blue (red), following a simple qualitative rule: if the initial conductance of the device is smaller (larger) than the immediate post-stimuli conductance then this event is considered as STP-F (STP-S). Initially, a steadily increase in the memristor's conductance is observed. When however a critically high conductance is reached, depressing-like events are activated to restore the low conductance level; this trend occurs consistently when observed over a long period of time.

We further studied the effect of the amplitude and rate of the stimulating scheme in controlling the short-term dynamics of our prototypes. Figure 3a shows the transient conductance change for a two-pulses (spikes) input for inter-spike intervals ranging from 20ms up to 200ms. This change is recoverable after a period of time as measurements after 1-120s show (see also supplementary material Figure S10). A clear correlation is found between the conductance decay and the interpulse timing, with the decay time constant being smaller for lower pulsing rates, as also illustrated in Figure 3a. This adheres with the notion that when repeated training of an event occurs within a short period of time it becomes more difficult to forget this event. Details about the fitting and parameter extraction methods are found in the supplementary material. Figure 3b depicts the contribution of each pulse stimulus to the device's conductance as a function of the stimulus amplitude; large amplitudes contribute a higher conductance modulation. The volatile behavior of our prototypes can be reproduced by an equivalent SPICE circuit model that we have presented previously[61], as demonstrated for example in Figure 3c.

## Exploiting metastable switching dynamics for processing spatio-temporal spike patterns

Both the short- and long-term plasticity mechanisms that characterize the ReRAM memristors described in this work have temporal properties and dynamics that are well within the range of biological cortical synapses. In addition to being an extremely useful property for directly emulating the properties of real synapses (e.g., for bio-hybrid systems and basic neuroscience research), this feature is very appealing for neuromorphic electronic systems. Neuromorphic systems comprise large arrays of neural processing elements in which memory and computation are co-localized, and in which time represents itself: the synapse and neuron circuits in these architectures process input spikes as they arrive, and produce output responses in real-time. Consequently, in order to interact with the environment and process real-world sensory signals efficiently, these systems must use computing elements that have biologically plausible time constants. By combining the advantages of ReRAM memristors with the properties of subthreshold analog neuromorphic VLSI circuits, it is therefore possible to build extremely compact and low-power neural processing systems that can interact with the environment in real-time[62]. In [62] the authors demonstrated a hybrid CMOS/memristor circuit that can use the conductance changes of ReRAM memristors to produce post-synaptic currents that have dynamics and properties very similar to the ones measured from real synapses (e.g. in terms of current amplitudes, time constants, etc.) When interfaced to neuromorphic silicon neuron circuits[63] one can implement efficient neural processing systems.

For example, one of the basic requirements of neural processing systems is their ability to recognize different patterns encoded in the temporal sequences of spikes produced by multiple neurons, e.g. at the sensory periphery. It is essential therefore to be able to distinguish different sets of spatio-temporal spike patterns quickly and efficiently. Here we show, perhaps, the first example of how memristors can be used to achieve spatiotemporal computation by performing an experiment with a static resistor and a memristor connected to a circuit that implements an exponential IF neuron model[64] (see also supplementary material, Figures S4-7 ). In Figure 4a we show the neural network diagram of the circuit designed to discriminate between two spatio-temporal patterns: the first pattern is represented by the sequence of events AB, when the spike-train labeled A appears before the one labeled B; the second pattern is represented by the sequence BA when the spike trains are sent in reversed order. The spike trains consist of three -4V, 10μs wide pulses with inter-spike interval $t_{int}$=250ms. For the pattern AB, the spike train is applied first to the static ($R_S$) and then to the dynamic (M) synapse, as illustrated in Figure 4b. The same train of pulses, but in reversed order, is applied in the case of the pattern BA as in Figure 4f. This sequence detector was designed to produce a spike at the membrane of the IF neuron when the pattern BA occurs, based on the short-term facilitating response of the memristor, and no spike when pattern AB occurs. The experiment was conducted by first applying five AB patterns, followed by five BA patterns and then repeating this full sequence eight times.

The success rate of the discrimination task is 67.5% with 15% false positive, as illustrated in Figure 4h. It was found that the variability in success is a cumulative effect of the response of the memristive synapse and the inherent noise of the system coupled with a low membrane voltage threshold. The performance of this network is compatible with the fact that biological neurons are inherently unreliable. Achieving reliability with unreliable circuits via redundancy strategies is a well-established concept in nature[65,66] that can be exploited also in these types of applications. A control experiment was performed in a similar fashion where the memristor was replaced by a static resistor in parallel with a capacitor (Figure S11), showing that in the absence of short-term facilitation (menristive synapse) we are unable to reliably discriminate between the events AB and BA (see Figure 4h).

The occurrence of false negatives, which in turn render the spiking probability of the neuron to be less than 100% is due to the probabilistic response of our DUT being facilitating (Figure 2d) or saturating (Figure 2e), with Figure 5c illustrating such a case. As the contribution to the membrane potential decreases with each subsequent input pulse, the membrane potential does not reach the pre-set threshold, prohibiting the neuron from spiking, as depicted in Figure 5b. It is however interesting to see that the opposite also holds; we have measured false positive cases where the neuron erroneously spikes when presented with an AB event. This is represented in Figures 5d,e,f and is again due to the synapse exhibiting an STP-S response rather than STP-F. Our proof-of-concept example of memristor-based sequence detectors, as presented in Figure 4, is only a simple scenario for highlighting the potential of this application. This concept is clearly amenable both to up-scaling, where multiple memristors and/or more complicated spatio-temporal patterns are employed, as well as the use of different neuronal models or even circuit parameters (e.g. membrane potential thresholds); opening remarkable opportunities for advancing the complexity of this system. Figure S12 demonstrates the same concept with aid of two volatile memristive synapses, illustrating a coincidence detector circuit.

## Summary

In this work we presented detailed and quantitative parallels between memristive devices and biophysically realistic models of synaptic dynamics. In particular we showed how meta-stable memory transitions that are typically seen as non-ideal effects that contribute to the large variability observed in emerging ReRAM, are in fact a salient feature for the establishment of truly biomimetic synapses that can faithfully reproduce short-term synaptic dynamics. In addition to reproducing plasticity mechanisms at a phenomenological level, these devices can be biased to exhibit both stochastic properties and biologically plausible temporal dynamics. These features can be exploited for developing non-von Neuman computing architectures in which memory and computation are co-localized, and where massively parallel circuits can process signals in real-time. Their ability to implement biologically realistic time constants would enable the construction of neural computing systems that can efficiently process real-world biologically relevant sensory signals and interact with the environment. Encompassing all these features in a single compact low-power device offers enormous potential for the development of real-time neuromorphic computing systems. Our approach paves the way towards performing a Turing-test for neuromorphic engineering, where emerging synaptic mimics are tested as in electrophysiology experiments and fitted by biophysically realistic models, while experts are invited to comment on whether the exhibited response stems from a real or an artificial system.

**Figure 1** Solid-state $TiO_2$ ReRAM memristors can support volatile switching, enabling the emulation of short-term plasticity. A top-view of a $2x2\mu m^2$ active area and 10nm thick $TiO_2$ cross-bar architecture is shown in a), with insets I and II respectively depicting cross-sections of a chemical synapse and a pristine memristor (blue denotes the Pt TE and BE that correspond to pre- and post-synaptic terminals, with green and red corresponding to Ti and $O_2$ species that can be displaced within the functional core). Schematic illustrations of possible conduction mechanisms that can induce a non-volatile increase (decrease) in the device's conductance, corresponding to an LTP (LTD) response, as depicted in and b); metastable transitions precede both cases, with potentiation (STP) captured respectively. A transition from volatile to non-volatile programming, as described in b), is shown in c) where identical voltage pulses initially trigger volatile switching and eventually a non-volatile state transition.

**Figure 2** A single memristor functioning both as a facilitating and saturated synapse. Shown are: a) repeated STP-F and b) STP-S post-synaptic response with simulation of the contributions of each pulse to a pre-synaptic neuron's membrane potential with τ=50ms and appropriate STP model fitting (details in supplementary materials), c) illustrates the pre-synaptic pulsing sequence applied to the memristive synaptic mimic; d) and e) delineate the corresponding occurrence probability of STP-F and STP-S events with respect to $G_0$, while f) presents measured transient conductance drift invoked by a train of three pulses (width=10μs, $t_{int}$=200ms, $t_{rec}$=20s), repeated 20 times.

**Figure 3** Device volatility controlled via timing and amplitude of stimuli. Shown are: a) memory-state decay measured after stimulus by a two-pulse train of 4V, 10μs wide and with interpulse time ranging from 20ms up to 200ms, b) correlation between normalized conductance changes induced in the memristor with respect to the amplitude of a 10μs width pulse and c) measured and simulated an STP-F event by employing an empirical PSPICE memristor model[61] (simulated and experimental results are indicated via a thick red and a thin blue line respectively).

**Figure 4** Demonstration of short-term dynamics in detecting concurrency of events. a) Illustration of the sequence detector circuit b) Pulsing sequence for Event AB – Event A is applied on the static synapse (resistor) while Event B is applied on the memristor M. c) Measured transient response of the neuron membrane potential for Event AB. d) Measured short-term dynamics of the memristor during Event AB that caused the neuron response shown in c) and averaged response of all AB events. e) Pulsing sequence for Event BA. f) Measured transient response of the neuron membrane potential for Event BA. g) Measured short-term dynamics of the memristor during Event BA that caused the neuron response shown in f) and averaged response of all BA events. h) Neuron spiking probability for Event AB and BA, benchmarked against a control experiment (details appear in Figure S11).

**Figure 5** Measured false positive and false negative cases. Shown are: a) event BA pulsing encoding, b) measured neuron membrane potential and c) transient conductance response of the memristive synapse following an STP-S trend, d) event AB pulsing encoding, e) measured neuron membrane potential and f) transient conductance response of the memristive synapse following an STP-F trend.

**Methods Summary**

All device prototypes exploited in this work were fabricated by the following process flow. 200nm of $SiO_2$ was thermally grown on top of 4-inch Si wafer, with 5nm Ti and 30nm Pt layers deposited via electron-gun evaporation to serve as the bottom electrodes (Ti is used as an adhesion layer). An RF magnetron sputtering system was used to deposit the active $TiO_2$ core from a stoichiometric target, with 30sccm Ar flow at a chamber pressure of P=$10^{-5}$mbar. Finally, all top Pt electrodes were deposited by electron-gun evaporation. A lift-off process was employed for patterning purposes prior each metal deposition. Good lift-off was accomplished via using two photoresist layers, LOR10 and AZ 5214E respectively, and conventional contact optical photolithography methods were used to define all layers. All finalized wafers were then diced, to attain 5x5mm$^2$ memristor chips, which were wire-bonded in standard packages for measurements. Preliminary characterization of all samples took place on wafer by employing a Wentworth semi-automatic prober and a Keithley SCS-4200 semiconductor characterization suite.

The cross-section of our memristor prototypes appearing on the inset of Figure 1a is a 256x256 pixel EDX map of a pristine (as-fabricated) device. This map was taken at 50μs dwell time, 1.2nA beam current and 8mins acquisition time on a FEI Titan G2 ChemiSTEM 80-200 microscope.

**Supplementary Information** is available in the online version of the paper.

**Acknowledgements** We acknowledge the financial support of the eFutures XD EFXD12003-4, the CHIST-ERA ERA-Net and EPSRC EP/J00801X/1, EP/K017829/1 and FP7-RAMP.

**Author Information** Reprints and permissions information is available at. The authors declare no competing financial interests. Correspondence and requests for materials should be addressed to T.P. (t.prodromakis@soton.ac.uk).

**Figure 1**

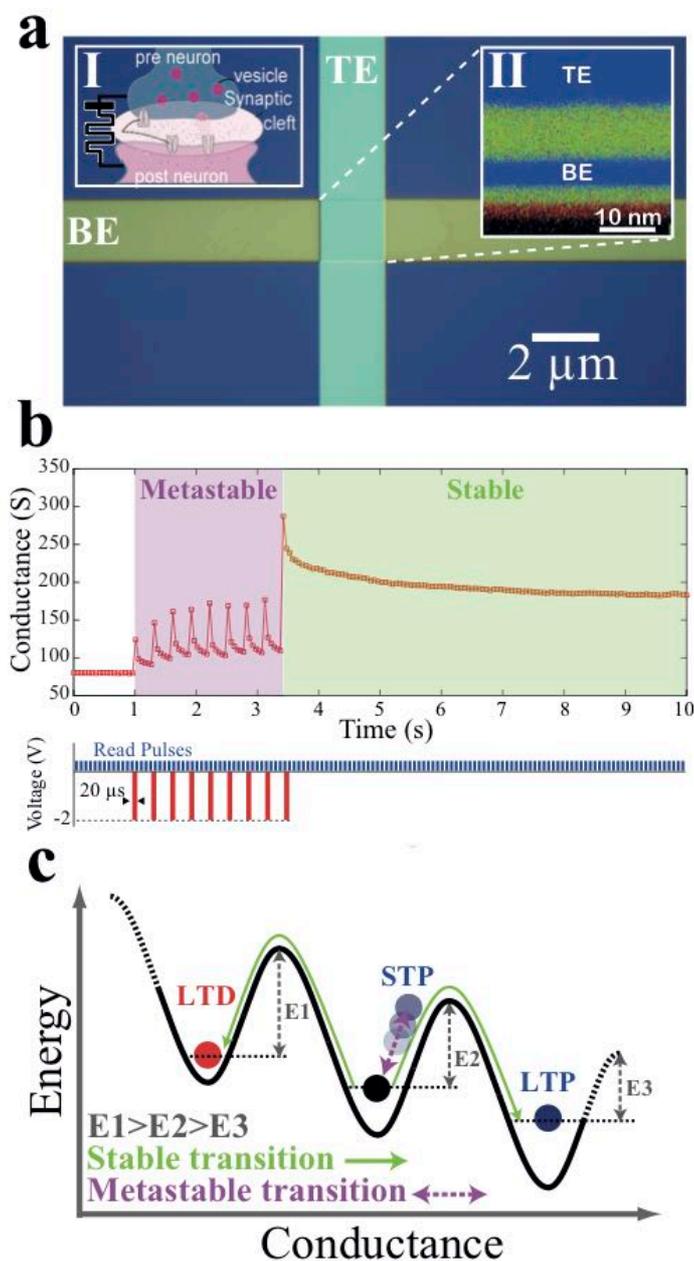

**Figure 2**

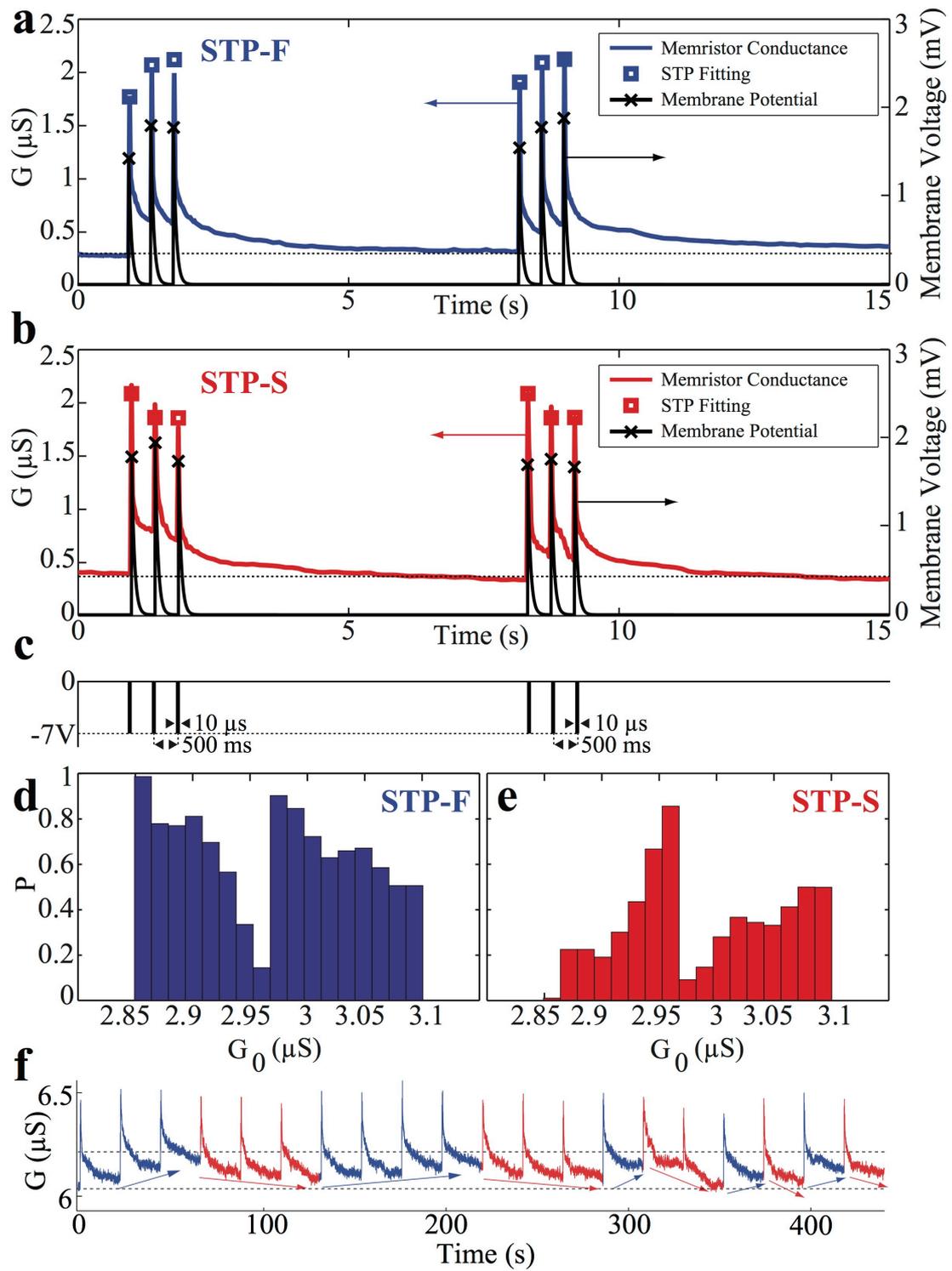

**Figure 3**

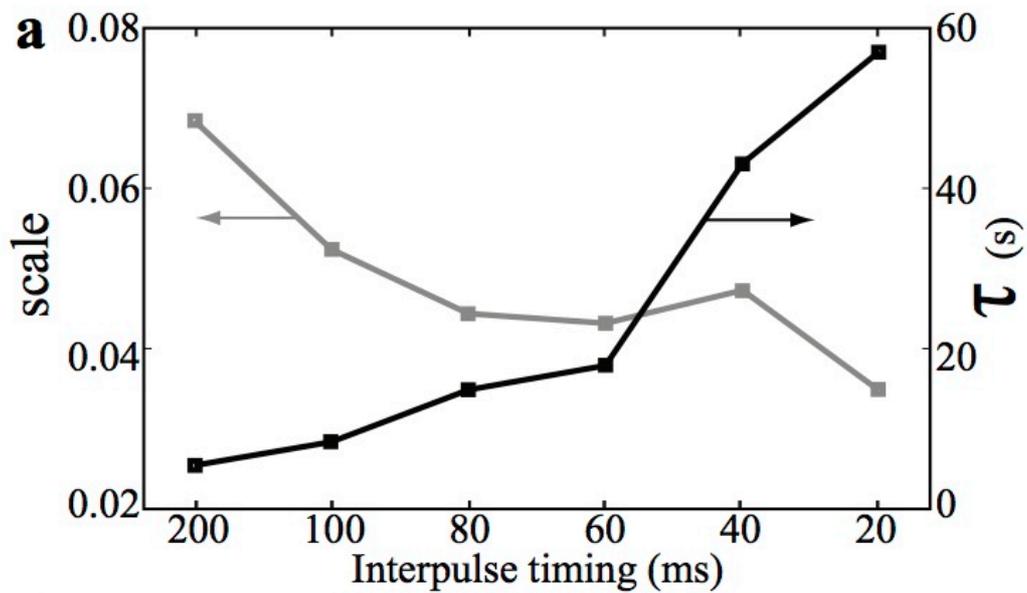

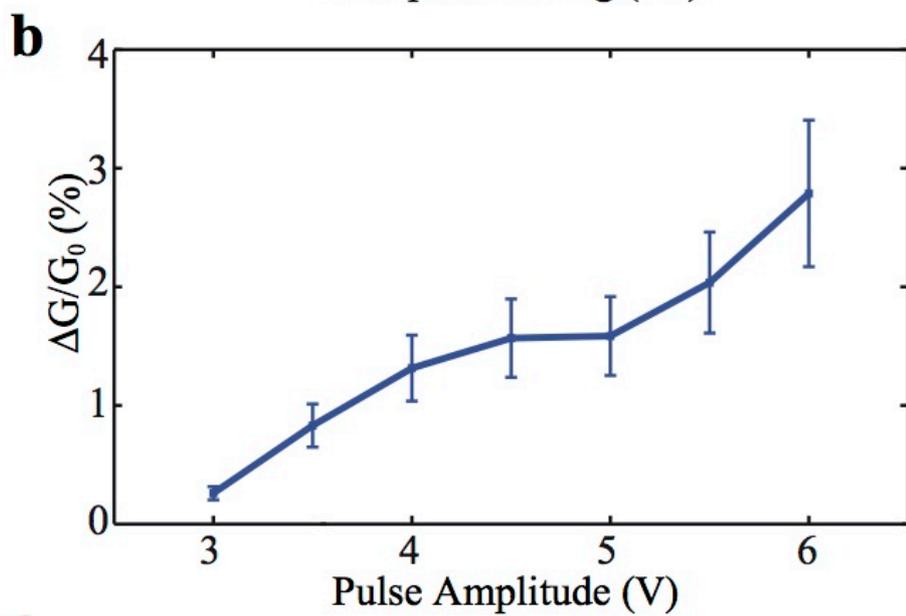

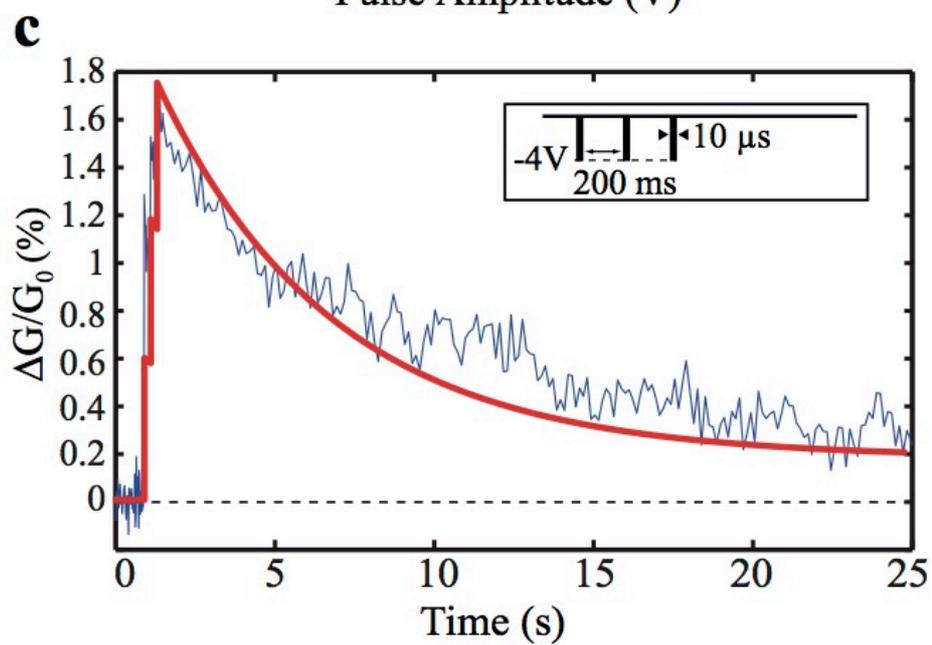

**Figure 4**

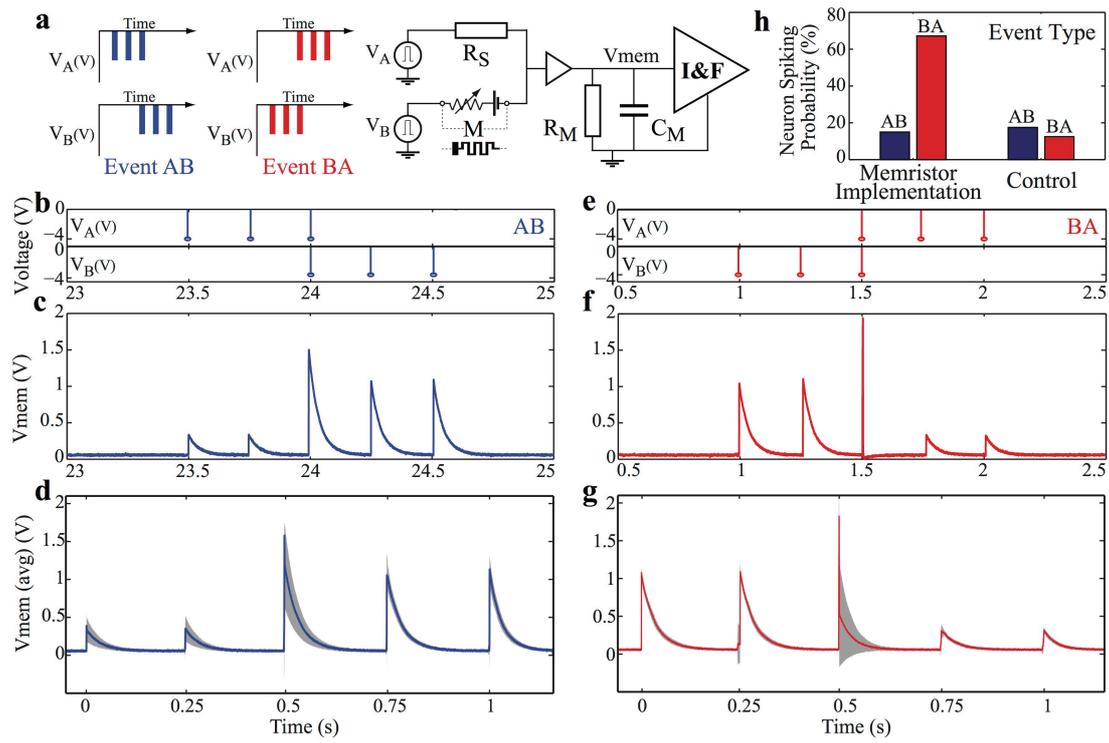

**Figure 5**

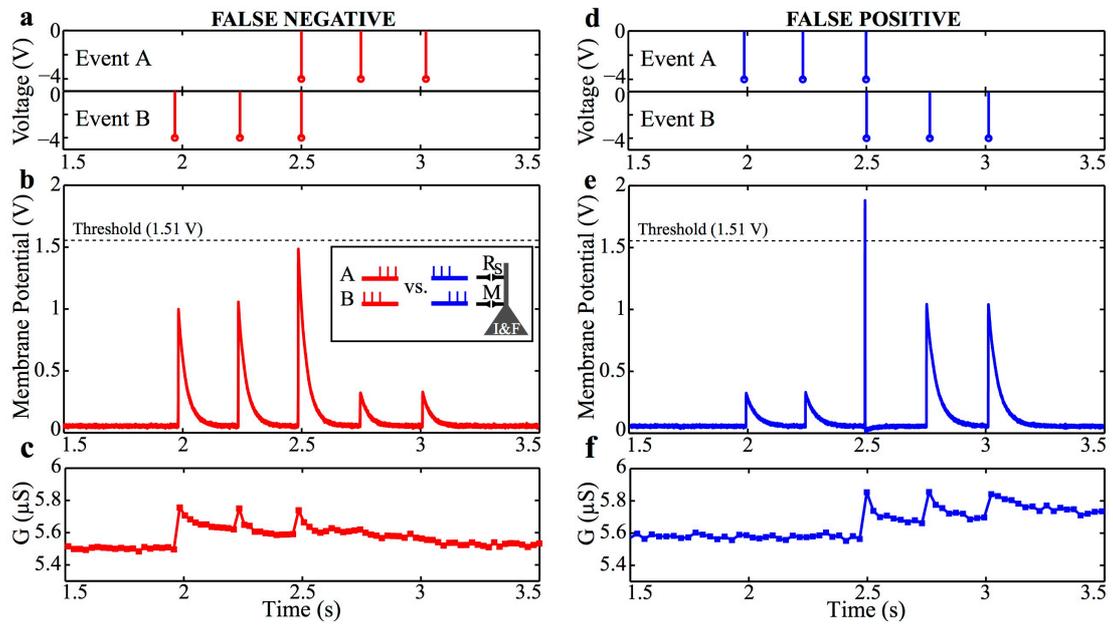